\documentclass[proof,mathleft]{WileyASNA-v1}

\articletype{Original Article}%

\received{26 April 2016}
\revised{6 June 2016}
\accepted{6 June 2016}

\usepackage{graphicx}
\usepackage{times}
\usepackage{hyperref}

\hypersetup{
  hidelinks,
  colorlinks=false,
}

\usepackage{times}

\usepackage{natbib}
\bibpunct{(}{)}{;}{a}{}{,}

\renewcommand{\vec}[1]{\mbox{\boldmath $#1$}}

\begin{document}

\title{Alpha tensor and dynamo excitation in turbulent fluids with anisotropic conductivity  fluctuations}

\author[1,2]{Oliver Gressel}

\author[1,3]{G{\"u}nther R{\"u}diger\,*}

\author[1]{Detlef Elstner}

\authormark{Gressel, R{\"u}diger \& Elstner} % \textsc{et al}}

\address[1]{\orgdiv{MHD \& Turbulence Section},\newline \orgname{Leibniz Institute for Astrophysics Potsdam}, \orgaddress{An der Sternwarte 16,\newline 14482
Potsdam, \country{Germany}\medskip}}

\address[2]{\orgdiv{Niels Bohr International Academy},\newline \orgname{The Niels Bohr Institute},\newline \orgaddress{Blegdamsvej 17, \newline DK-2100, Copenhagen \O, \country{Denmark}\medskip}}

\address[3]{\orgdiv{Institute of Physics and Astronomy}, \orgname{University of Potsdam}, \orgaddress{Karl-Liebknecht-Str. 24-25,\newline 14476 Potsdam, \country{Germany}}}

\corres{*\,\email{gruediger@aip.de}}

%\presentaddress{This is sample for present address text this is sample for present address text}

\abstract{A mean-field theory of the electrodynamics of a turbulent fluid is formulated under the assumption that the molecular electric conductivity is correlated with the turbulent velocity fluctuation in the (radial) direction, $\vec g$.  It is shown that for such homogeneous fluids a strong turbulence-induced field advection anti-parallel to $\vec g$ arises almost independently of rotation. For rotating fluids, an extra $\alpha$\,effect appears with the known symmetries and with the expected maximum at the poles. Fast rotation, however, with Coriolis number exceeding unity suppresses this term.  Numerical simulations of forced turbulence using the {\sc nirvana} code demonstrate that the radial advection velocity, $\gamma$, always dominates the $\alpha$\,term. We show finally with simplified models that $\alpha^2$\,dynamos are strongly influenced by the radial pumping: for $\gamma<\alpha$ the solutions become oscillatory, while for $\gamma>\alpha$ they become highly exotic if they exist at all. In conclusion, dynamo models for slow and fast solid-body rotation on the basis of finite conductivity--velocity correlations are unlikely to work, at least for $\alpha^2{\it \Omega}$~dynamos without strong shear.\bigskip}

\keywords{astrophysical plasma -- dynamo theory}

\jnlcitation{\cname{%
\author{O. Gressel}
\author{G. R{\"u}diger}, and
\author{D. Elstner}}
(\cyear{2021}),
\ctitle{Alpha tensor and dynamo excitation in turbulent fluids with anisotropic conductivity fluctuations}, \cjournal{Astronomical Notes}, \cvol{2021;00:x--y}.}

%%\fundingInfo{Funding info text.}

\maketitle

% -------------------------------------------------------------------------------

% my things
\def\qq{\qquad\qquad}
\def\qqq{\qquad\qquad\qquad}

\def\d{{\rm d}}
\def \i{{\rm i}}
\def \n   {\vec{\nabla}\!}
\def \p   {\partial}
\def \t   {\times}
\def \Om  {{\it \Omega}}
\def \rin {r_{\rm in}}
\def \mur {r_{\rm in}}
\def \Rey {\ensuremath{\rm{Re}}}
\def \Ha {\ensuremath{\rm{Ha}}}
\def \Hamin {\ensuremath{\rm{Ha_{min}}}}
\def \Pm {\ensuremath{\rm{Pm}}}
\def \Rm {\ensuremath{\rm{Rm}}}
\def \Mm {\ensuremath{\rm{Mm}}}
\def \S  {\ensuremath{\rm{S}}}
\def \Ru {\ensuremath{\overline{\rm Rm}}}
\def\beg{\begin{equation}}
\def\ende{\end{equation}}
\newcommand{\gsim}{\lower.7ex\hbox{$\;\stackrel{\textstyle>}{\sim}\;$}}
\newcommand{\lsim}{\lower.7ex\hbox{$\;\stackrel{\textstyle<}{\sim}\;$}}
\renewcommand{\vec}[1]{\mbox{\boldmath $#1$}}
\def\curl{{\rm curl}}
\def\div{{\rm div}}

\def\Om{{\it \Omega}}
\def\R{{R\"udiger}}
\def\A{{Alfv\'en}}
\def\etaT{\eta_{\rm t}}
\def\meta{{\bar \eta}}
\def\etarms{\eta_{\rm rms}}
\def\urms{u_{\rm rms}}
\def\nat{Nature\ }
\def\aa{ Astronomy \& Astrophysics\ }
\def\aap{ Astronomy \& Astrophysics\ }
\def\aas{ Astronomy \& Astrophysics Suppl.}
\def\ara\&a{ Ann. Rev. Astronomy Astrophysics}
\def\aarev{ Astronomy \& Astrophysics Rev.}
\def\apj{ The Astrophysical Journal\ }
\def\apjl{ The Astrophysical Journal Letters\ }
\def\apjs{ Astrophys. J. Suppl.}
\def\apss{ Astrophys. Space Sci.}
\def\aj{ Astronomical Journal}
\def\jfm{ J. Fluid Mechanics\ }
\def\arafm{Ann. Rev. Fluid Mechanics\ }
\def\gafd{Geophysical Astrophysical Fluid Dyn.\ }
\def\mnras{Month. Not. Roy. Astr. Soc.\ }
\def\an{ Astronomische Nachrichten\ }
\def\sp{ Solar Physics\ }
\def\grl{ Geophysical Research Letters\ }
\def\jgr{ J. Geophysical Research\ }
\def\prl{Physical Review Letters\ }
\def\pre{ Physical Review E\ }
\def\prb{ Physical Review B\ }
\def\ssr{ Space Science Review\ }
\sloppy

% -------------------------------------------------------------------------------

\section{Introduction}

% -------------------------------------------------------------------------------

If apart from the velocity and magnetic field, by any reason also the electric conductivity in a turbulent fluid fluctuates around a certain value then also the local magnetic diffusivity fluctuates around its average.  \cite{KR73} started to consider the consequences of this constellation with the result that the effective decay time of a large-scale nonuniform magnetic field is changed by reducing the effective eddy diffusivity of the turbulence field.

Moreover, in convection-driven turbulent fields the always existing temperature fluctuations should produce magnetic resistivity fluctuations which are correlated with one of the velocity components, e.g., the vertical one. In this case, even a turbulent diffusivity-flux vector $\langle\eta'\vec{u}'\rangle$ ---with $\eta=1/\mu_0\sigma$ denoting the magnetic resistivity and $\vec u'$ the velocity fluctuations--- occurs.  This, in connection with the magnetic background field or electric current, may form new terms in the mean-field induction equation. \cite{PA16} suggested a new sort of $\alpha$\,effect arising in such systems.

They derived an expression for the diffusivity-current correlation, in which the diffusivity-flux vector, multiplied with the mean magnetic field, $\bar{\vec B}$, appears so that a new $\alpha$~effect could be possible in spite of the assumed homogeneity of the turbulence field. However, there are two possibilities for the relation between the electromotive force and the mean magnetic field: the latter can be ~i) parallel to the electromotive force or ~ii) perpendicular to the electromotive force. Only in the first case, one formally speaks of an $\alpha$~effect, which may lead to self-excitation of large-scale magnetic fields, while in the second case the expression describes a turbulent diamagnetism (also called ``topological pumping'') which is known to hamper dynamo instability.  If the correlation $\langle\eta'\vec{u}'\rangle$ exclusively defines a preferred direction $\vec g$ the resulting turbulent electromotive force is perpendicular to the mean magnetic field and an alpha-effect is not obtained.

Later on, quasi-linear SOCA calculations applicable to rotating forced turbulence and/or magneto-convection indeed confirmed the existence of an $\alpha$\,effect in the presence of global rotation. Without rotation, the conductivity fluctuations lead to a reduction of the eddy diffusivity and ---if correlated with one of the velocity components--- to a new but rather strong diamagnetic pumping effect \citep{RK20}. In that work, rotating magneto-convection was numerically used to derive the radial turbulent electric current flux $\langle u'_r\curl\vec{B'}\rangle$ ---where $r$ is the radial coordinate--- which serves as a proxy of the turbulent diffusivity-current vector $\langle\eta'\curl\vec{B'}\rangle$ if $\eta'$ and $u'_r$ are correlated or anti-correlated. The flux vector always exists for rotating convection under the influence of an azimuthal magnetic background field. The result is a well-defined diamagnetic pumping and, with rotation, an $\alpha$\,effect which is anti-symmetric with respect to the equator.

However, convection only exists if the fluid is stratified in the radial direction, $\vec g$. The main difference caused by the fluctuating-conductivity concept is the occurrence of an $\alpha$\,effect in fully uniform fluids in which an anisotropy exists rather than any form of stratification. This makes the idea a promising one for a dynamo theory of planetary magnetism.

In the present paper, therefore, the existence of the $\alpha$\,effect in absolutely homogeneous fluids is shown by numerical simulations of forced rotating turbulence. We shall demonstrate that the $\alpha$\,effect indeed occurs, if the global rotation is not too slow or too fast but that it is, however, always accompanied by a dominating diamagnetic pumping term, $\gamma$. Even without rotation (and only slightly suppressed in its presence) a strong radial advection term occurs by which the horizontal field (i.e. perpendicular to $\vec g$) is lifted to either of the radial boundary layers, depending on the sign of the effect.

We note that a large-scale $\alpha^2$\,dynamo can in principle operate for very {\it weak} $\alpha$\,effect if only the region is big enough, or ---with other words--- if it hosts a sufficiently large number of eddies. In our final Section, the consequences of this puzzling situation are shown by the presentation of a sequence of mean-field $\alpha^2$\,dynamo models with stronger and stronger magnetic pumping term (i.e. turbulence-induced diamagnetism). We shall show that such dynamos can only operate as long as the $\alpha$\,term (in form of a pattern velocity) exceeds the pumping velocity. This condition is unfortunately not met -- at least, according to the results of the derived electrodynamics, which is based on the correlations with conductivity fluctuations.

% -------------------------------------------------------------------------------

\section{The Equations}
\label{sec:Equs}

% -------------------------------------------------------------------------------

The basic   equation of the problem is the induction equation
\begin{eqnarray}
 \frac{\partial \vec{B}}{\partial t}&=& {\textrm{curl}} \Big(\vec{u} \times \vec{B} -
 \eta\ {\textrm{curl}} \vec{B}   \Big)\,,
   \label{mhd1}
\end{eqnarray}
with the continuity condition $\,{\textrm{div}}\,\vec{B} = 0$. Moreover, we assume $\,{\textrm{div}}\,\vec{u} = 0$ as the condition for an incompressible fluid for the analytic derivations, while for the numerical experiments, this constraint is relaxed.
Here, $\vec{u}$ is the fluid velocity, $\vec{B}$ is the magnetic field
vector and $\eta$  the (molecular) magnetic diffusivity. We consider a turbulent fluid with
$\vec{u}= \bar{\vec{u}}+ \vec{u}' $ and with a fluctuating magnetic
diffusivity ${\eta}= \meta+ {\eta}' $. For the expectation values of
the perturbations we shall use the notations
$\urms={\langle {\vec{u}'}^2 \rangle}$$^{1/2}$ and $\etarms={\langle {\eta'}^2 \rangle}$$^{1/2}$.
{ Large-scale observables { (i.e., mean values)} are marked with overbars,
 while brackets are used for the correlations {of fluctuations}.
 Low or high  values of the  magnetic Reynolds number
\beg
{\Rm}=\urms\ell/\meta
\label{Rm}
\ende
(for Strouhal number $\simeq 1$, and with $\ell$ the correlation length) distinguish between the regimes of low\,/\,high conductivity.
Within the realm of the electrodynamics with finite fluctuations,
the high-conductivity limit $\meta\!\to\! 0$ may not be allowed.

If the fluctuations $\vec{u}'$ and $\eta'$ exist and are correlated, then the
 turbulence-originated diffusivity flux
 \beg
\vec{U}=\langle \eta' \vec{u}'\rangle
\label{U}
\ende
forms a vector, which is polar  by definition. The existence of the radial component of this vector is obvious for thermal convection, where both the radial velocity and the electric conductivity are due to temperature fluctuations. The correlation (\ref{U}) can be understood as  transport of magnetic diffusivity in a  certain  direction. If, e.g., the correlation between $\eta'$ and $u_r'$ is positive then resistivity is transported upwards -- balanced by a downward radial velocity $\nabla(- \eta)$ which ``pumps''  the horizontal  field downwards in the direction where the magnetic decay is maximum (the ``diamagnetic effect'' of turbulent origin).

Also the magnetic field will fluctuate, hence $\vec{B}\!=\!\bar{\vec{B}}+ \vec{B}' $. The magnetic fluctuation $\vec{B}'$ fulfils a nonlinear  induction equation which follows from (\ref{mhd1}).
The turbulence-originated electromotive force
${\vec{\cal E}}=\langle \vec{u}'\times \vec{B}' \rangle$ and the  diffusivity-current correlation
$
\vec{{\cal J}}=-\langle \eta' \curl
\vec{B}'\rangle
$
enter  the induction equation for large-scale magnetic field via
\beg
 \frac{\partial \bar{\vec{B}}}{\partial t}= {\textrm{curl}} \Big(\vec{\cal E} + \vec{\cal J}
- \meta\ {\textrm{curl}} \bar{\vec{B}}   \Big)\,.
   \label{mhd0}
\ende
Under the assumption that the large-scale  field, $\bar{\vec{B}}$,   varies sufficiently slowly in space and time, the electromotive force can be written as
\beg
\vec{{\cal E}} = \alpha \circ \vec{\bar{B}} - \etaT \curl \vec{\bar{B}}\,,
\ende
where the tensor $\alpha$ and the coefficient $\etaT$ represent the $\alpha$\,effect and the turbulent magnetic diffusivity \citep{KR80}, respectively, and where `$\circ$' denotes a tensor multiplication. The tensorial structure of $\etaT$ under the presence of magnetic field and rotation has been discussed later by \cite{KP94}.  As in \cite{RK20}, the spectral vector of the correlation (\ref{U}) may be written as
\beg {\hat U}_i=u_1(k,\omega) \left( g_i-\frac{(\vec{g}\!\cdot\!\vec{k})\,k_i}{k^2} \right) .
\label{Ucorr}
\ende
The vector $\vec g$  gives the unit vector of the
direction in which the correlation between velocity and diffusivity is
non-vanishing. The expression (\ref{Ucorr}) must be odd in $\vec{g}$ and its real
part must be even in the wave number $\vec{k}$. The quantity $u_1$
reflects the correlation of the velocity component $\vec{g}\!\cdot\!\vec{u}'$
with $\eta'$ where $\omega$ is the Fourier frequency of the spectrum. As it should, the transformation $\vec g \to -\vec g$ only changes the sign of $\vec U$.

% -------------------------------------------------------------------------------

\section{The  diffusivity-current correlation}

% -------------------------------------------------------------------------------

It has been shown earlier  that a relation
\beg
 \vec{{\cal J}} =-\gamma \ \vec{g}\times \bar{\vec{B}}
\label{dia}
\ende
between the the diffusivity-current correlation, $\vec{{\cal J}}$, and the large-scale magnetic field, $\bar{\vec{B}}$, results with
\beg
 \gamma= \frac{1}{3} \int\!\!\!\!\int \frac{\meta k^4  u_1}{{ \omega^2 + \meta^2 k^4} }\; {\rm d}\vec{k}\, {\rm d} \omega\,,
\label{gamma}
\ende
representing a turbulent advection of the magnetic background field  where ${\vec u}_{\rm adv}=-\gamma \vec{g}$ is the advection velocity   \citep{RK20}. We find a coefficient $\gamma$ of the same sign as the diffusivity flux (\ref{U}). For positive $u_1$ (i.e., for positive correlation of $\eta'$ and $u'_r$), the advection velocity, ${\vec u}_{\rm adv}$, points downward if  $\vec g$ is the radial unit vector.  Anti-correlated $\eta'$ and $u_r'$ lead to an upward turbulent transport of the mean magnetic field. This means that the field is always attracted by  the islands of lower resistivity -- or, equivalently, of higher electric  conductivity. As a consequence, the large-scale magnetic field favours the direction towards longer diffusive decay times. The advection velocity is opposite to the  diffusivity flux (\ref{U}).
The integral expression for $\gamma$ of Eq.~(\ref{gamma}) scales linearly with $\Rm$ until it saturates for large magnetic Reynolds numbers.

 Let $\hat V$ be the spectral function of the two-point autocorrelation function
$
V(\vec{\xi}, \tau)=\langle \eta'(\vec{x},t)\,\eta'(\vec{x}+\vec{\xi}, t+\tau)\rangle
$
of the diffusivity fluctuations.
For the diffusivity-current correlation $\vec{\cal J}$ the term with  $\hat V$  leads to
\beg
\vec{\cal J}= \dots +  \frac{2}{3}
\int\!\!\!\!\int  \frac{k^2 {\hat V}}{-{\rm i} \omega + \meta k^2 }\;
{\rm d}\vec{k}\, {\rm d}\omega \ \curl \bar{\vec{B}}\,,
\label{mhd7}
\ende
which provides an extra contribution to the magnetic field
dissipation. The question is whether this term reduces or enhances the
  eddy diffusivity $\etaT$ representing turbulence without $\eta$-fluctuations.
The small-scale diffusivity fluctuations obviously  lead to a {\it reduction} of the large-scale eddy diffusivity $\etaT$ which, however, is only weak as it  runs with the small value $(\etarms/\bar\eta)$ in second order \citep{KR73,RK20}. The actual value of the turbulence dissipation
will not have relevance for the results of the present paper.

Our assumed background turbulence is homogeneous but anisotropic, where  the anisotropy is only implicit. If the turbulence rotates, an additional  pseudo-scalar $\vec{g}\cdot\vec{\Om}$ appears
with which a relation
 \begin{eqnarray}
\vec{\cal J}&=&
 -\gamma\ \vec{g}\times \vec{\bar B}-\nonumber\\ &-&\alpha_1 \left[\; (\vec g\!\cdot\! \vec{\bar B})\,\vec{\Om} +(\vec{g}\!\cdot\!\vec{\Om})\,\vec{\bar B}  \;\right] \;- \alpha_2  (\vec{\bar B}\!\cdot\!\vec\Om)\,\vec g
\label{J2}
\end{eqnarray}
can be formulated -- with yet unknown coefficients $\alpha_1$ and $\alpha_2$ for the diffusivity-current correlation, $\vec{\cal J}$, in presence of a large-scale magnetic field and rotation. For the above expression, $\gamma$ is again given by Eq.~(\ref{gamma}). Relation (\ref{J2}) formally describes the existence of an $\alpha$~tensor which connects the correlation $\vec{\cal J}$ with the large-scale magnetic field  $\vec{\bar B}$. This connection exists despite the turbulence model being assumed as strictly homogeneous (so that the standard $\alpha$ tensor cannot appear).
The $\alpha$\,effect according to (\ref{J2}) is highly anisotropic, the middle
term  with the coefficient $\alpha_1$ provides the rotation-induced standard $\alpha$ expression.
While the diamagnetic term with $\gamma$ also exists for $\Om=0$,  the $\alpha$ terms  need global rotation.  We shall  show below  that, independently of the sign of the correlations $\langle \eta' u'_r \rangle$,  the values of $\alpha_1$ and $\gamma$  are always of opposite sign.

\noindent The dimensionless ratio
\beg
\hat\gamma=\frac{\gamma}{\alpha_1\Om}
\label{gammadach}
\ende
 of the pumping velocity  $\gamma$ and the  rotation-induced $\alpha$
 effect indicates the ratio of anti-symmetric  and symmetric elements in the complete $\alpha$ tensor.
Simulating   electromotive forces for models of  rotating magnetoconvection,
\cite{OS01,OS02} found $\hat\gamma\simeq 1$  where both $\alpha$ and $\gamma$  were about 10\% of the rms value of the convective velocity.  Also \cite{KK09} reached  typical values of order unity in their numerical models  of turbulent magnetoconvection. Additionally, with their extensive numerical simulations,  \cite{GZ08}  derived $\hat\gamma = O(1)$ for interstellar turbulence driven by collective  supernova explosions. All these examples summarise the results of $\alpha$ effect calculations from the relation between the electromotive force $\vec{\cal E} $ and the mean magnetic field $\vec{\bar B}$, which only appears if the turbulence is nonuniform. On the other hand, we shall demonstrate in the following that for {\it homogeneous} models with fluctuating conductivities, the corresponding ratio (\ref{gammadach})  reaches values even exceeding unity -- with severe consequences for associated dynamo models.

% -------------------------------------------------------------------------------

\section{Numerical methods}

% -------------------------------------------------------------------------------

To probe the theoretical predictions we run artificially forced, fully nonlinear numerical simulations with the {\sc nirvana} MHD code \citep{Z04}, which solves the equations of compressible magnetohydrodynamics by means of a second-order Godunov approach.  In the simulations, the fluctuating component of the magnetic diffusivity is prescribed by
$
\eta' = c_u u_z,
$
where the coefficient $c_u$ is used to control the strength of the correlation. We furthermore use
$\eta_{\rm rms}=c_u u_{z, {\rm rms}}$
to quantify the
amplitude of the fluctuating part of the magnetic diffusivity.
The simulation domain is a fully periodic cube with volume $L^3$. The units of length and time are
$
[x] = k_1^{-1}, [t]=(c_{\rm s} k_1)^{-1}$
where $k_1$ is the wave number corresponding to the system size
and  $c_{\rm s}$ is the
constant speed of sound. The simulations employ standard non-helical forcing according to eqn.~(7) of \cite{HBD04} and are characterised by the  magnetic Reynolds number (\ref{Rm})
with $u_{\rm rms}$ volume averaged { and $\ell=(k_{\rm f})^{-1}$}.
The flows under
consideration are weakly compressible with Mach number ${\rm
  Ma}=u_{\rm rms}/c_{\rm s}\approx 0.1$. All simulations have
$k_{\rm f} \simeq 4.5$ (using isotropically sampled discrete wave vectors obeying $4 \le k_{\rm f} \le 5$) and employ a grid resolution of $80^3$. In code units, the molecular diffusivity is fixed at $\bar\eta=0.02$.

% -------------------------------------------------------------------------------

\section{The turbulent flux of electric current}

% -------------------------------------------------------------------------------

Consider a homogeneous and isotropic turbulence that is influenced by   uniform magnetic fields and global rotation.
 Let us write its correlation tensor, $\langle {u}_i' \  \curl_j \vec{B}'\rangle $,  as
\begin{eqnarray}
&\langle{u}_i'  &\curl_j \vec{B}'\rangle =\nonumber\\
&=&\kappa'  \epsilon_{jik} {\bar B}_k + \kappa_1 \Om_i {\bar B}_j +\kappa_2 \Om_j {\bar B}_i + \kappa_3 (\vec{\Om}\cdot \vec{\bar B})  \delta_{ij}\,.
\label{alpha4}
\end{eqnarray}
The tensor  is not a pseudo-tensor and  there is no reason that the dimensionless  coefficients $\kappa$ identically  vanish. It does not play a known role in the mean-field electrodynamics but it is exploited  here  as a proxy of the desired diffusivity-current correlation. The correlation   vector
$\langle {u}_r' \ \curl \vec{B}'\rangle$
  describes an upward  or downward radial   flux of  electric current  in a rotating magnetised turbulence which we shall  use below  to  estimate the diffusion-current correlation $\vec{\cal J}$.
 We note that  for $\Om=0$ it is
$
\langle (\vec{g}\cdot \vec{u'})\  \curl \vec{B}'\rangle =  \kappa' \vec{g} \times  \vec{\bar B}
$
for all directions $\vec{g}$. With $\vec g$ as the radial direction, one finds
\beg
\langle {u}_r'  \curl_\theta \vec{B'}\rangle= -\langle {u}_\theta'  \curl_r \vec{B'}\rangle = - \kappa' {{\bar B}_\phi}\,,
\label{kadash}
\ende
if the magnetic  background field  only has an azimuthal component. Based on SOCA calculations, the coefficient
$\kappa'$ is
\beg
\kappa'=\frac{1}{15}
\int\limits_0^\infty\!\!\int\limits_0^\infty  \frac{\eta k^4  E(k,\omega)}{ \omega^2 + \eta^2 k^4 }\;
{\rm d}{k}\, {\rm d}\omega\,,
\label{gamm6}
\ende
with  the positive spectral function $E$ of the turbulence intensity,
\beg
u^2_{\rm rms} =
\int\limits_0^\infty\!\!\int\limits_0^\infty E(k,\omega)\;{\rm d}k \ {\rm d}\omega\,.
\label{intensity}
\ende
As the spectrum $E(k,\omega)$ is  positive-definite, the tensor coefficient $\kappa'$ is positive-definite, too.

Figure~\ref{figkappa} gives  a numerical representation of the complete tensor (\ref{alpha4}) in Cartesian  coordinates $(r,\theta,\phi)\to(x,y,z)$ where the rotation vector is ${\vec \Om}=\Om_0(\cos\theta, -\sin\theta, 0)$  and the magnetic field  $\vec{\bar B}= (0,0,B_0)$. The details of the simulations were given in the previous Section. Obviously, the $\kappa_3$ coefficient in (\ref{alpha4})  cannot be determined for this geometry as always ${\vec\Om}\perp\vec{B}$. It is clear from the uppermost and the lowermost curves in the left and the right panel  that after (\ref{kadash}) the simulation gives $\kappa'>0$  in accordance to the result (\ref{gamm6}) of the quasi-linear theory. Only the $xy$-component is anti-symmetric in its indices but the  cross correlations $xz$ and $yz$ are symmetric. The diagonal components $xx$, $yy$ and $zz$ vanish (not shown) in accordance to the relation (\ref{alpha4}).

 For the remaining off-diagonal tensor components, one finds $\kappa_1=\kappa_2=\kappa$ with $\kappa<0$ as
\beg
\langle {u}_r'  \curl_\phi \vec{B'}\rangle= \langle {u}_\phi'  \curl_r \vec{B'}\rangle =  \kappa \Om {{\bar B}_0}\cos\theta<0
\label{kappass}
\ende
and
\beg
\langle {u}_\theta'  \curl_\phi \vec{B'}\rangle= \langle {u}_\phi'  \curl_\theta \vec{B'}\rangle = - \kappa \Om {{\bar B}_0}\sin\theta>0\,,
\label{kappas}
\ende
hence  for  rotating and magnetised  (but otherwise isotropic) turbulence,   the tensor expression (\ref{alpha4}) becomes
\begin{eqnarray}
\langle {u}_i'\  \curl_j \vec{B}'\rangle &=&\kappa'  \epsilon_{jik} {\bar B}_k +\nonumber\\ &+&\kappa ( \Om_i {\bar B}_j + \Om_j {\bar B}_i) + \kappa_3 (\vec{\Om}\cdot {\vec{\bar B}})  \delta_{ij}\,.
\label{alpha5}
\end{eqnarray}
In a rotating but otherwise isotropic turbulence with an azimuthal background field, the
meridional flow fluctuations  will always be  correlated with the azimuthal electric-current fluctuations.
We note that the simulations show that the anti-symmetric ($xy$)-component of the tensor is {\it always} much larger than the symmetric ($xz$)-component -- which, in fact, will have important consequences.

\begin{figure*}
  \centerline{
   \includegraphics[width=0.75\textwidth]{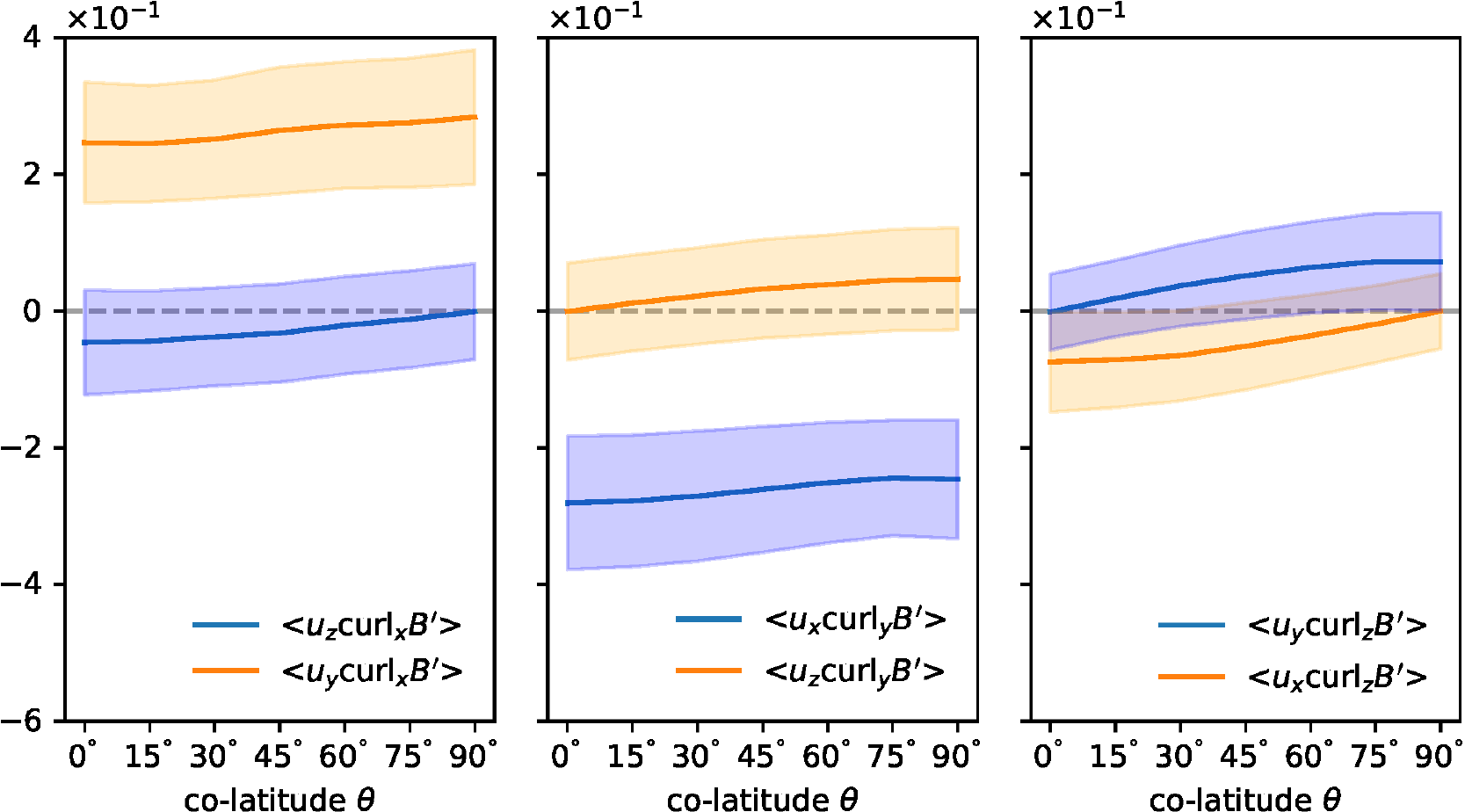}
   }
 \caption{The  off-diagonal components (expectation value plus temporal variations) of the turbulence-induced electric-current flux  tensor ${\langle u'_i\curl_j {\vec B}'  \rangle}$ normalised with $\urms k_{\rm f} B_0$ for various co-latitudes. The plot reflects the symmetry of the tensor except the $xy$-component which is anti-symmetric in accordance with Eq.~(\ref{alpha5}). $\Rm=100$, $B_0=2\times 10^{-8}$,  $\Pm=1$, $\Om=1$.}
\label{figkappa}
\end{figure*}

Replace now in the relations  (\ref{kadash}) and (\ref{kappass}) $u'_r$ by $\eta'$ and the existence of correlations such as   $\langle \eta' \curl_\theta {\vec B}'\rangle$ and $\langle \eta' \curl_\phi {\vec B}'\rangle$ becomes obvious in (rotating) homogeneous turbulence fields  magnetised with an azimuthal background field. Just this finding is formulated by Eq.~ (\ref{J2}). For positive correlation of the $\eta$-fluctuation and the radial velocities (i.e., positive $U_r$), the $\alpha_1$ in (\ref{J2}) becomes negative
and for negative correlations it becomes  positive.
Note  the negative sign in the definitions.
In the same relation, the $\gamma$ results as positive -- hence the pumping is downward (i.e., opposite to $U_r$). We always obtain $\gamma\alpha_1\leq 0$ for both signs of $U_r$.

Another basic finding is that the term with $\kappa'$
always exceeds those with $\kappa$, which -- in other words -- means that, for rotating turbulence, the pumping term (a velocity) will  always be larger than the  $\alpha$ term (also a  velocity).  As a consequence, in rotating  conducting fluids, the diamagnetic effect may by far exceed the inducting action of the $\alpha$ terms. The remainder of this paper will confirm this suggestion and will show that a dominating turbulent pumping precludes   dynamo  instability of the $\alpha^2$-type, that is, in the absence of large-scale shear.

\begin{figure*}
  \centerline{
 \includegraphics[width=0.75\textwidth]{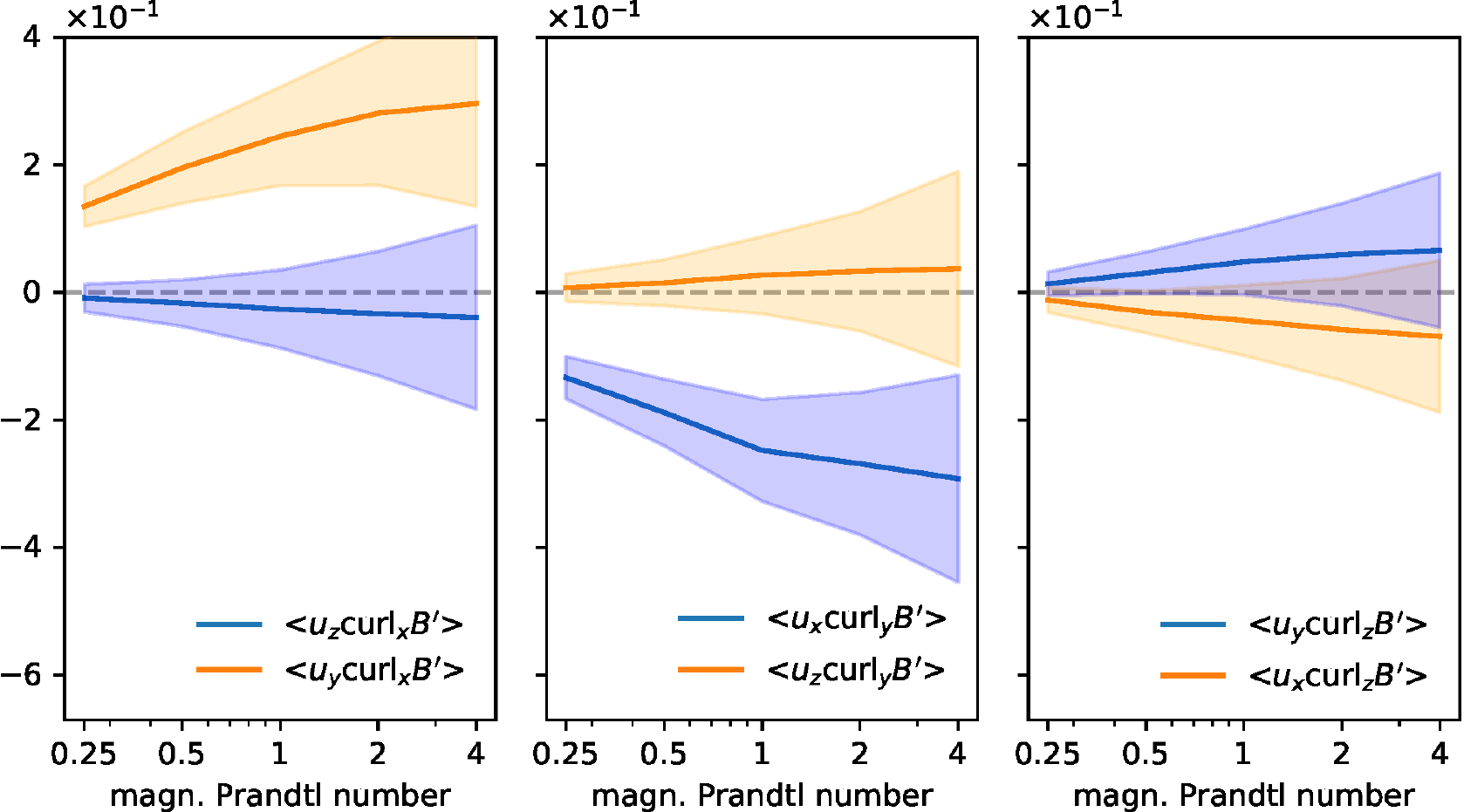}
 }
 \caption{Similar to Fig.~\ref{figkappa} but for $\theta=45^\circ$ and for various magnetic Prandtl numbers. The blue line in the middle panel ($\langle {u}_x'  \curl_y \vec{B'}\rangle$, leading to the  advection term) and the orange line in the right panel ($\langle {u}_x'  \curl_z \vec{B'}\rangle$, leading to the $\alpha$\,effect) are of particular  relevance. The  ratio $\langle {u}_x'  \curl_y \vec{B'}\rangle/\langle {u}_x'  \curl_z \vec{B'}\rangle$ for all $\Pm$  always exceeds unity.
 $\Rm=11$,  $\Om=1$.}
\label{fig4}
\end{figure*}

Figure \ref{fig4} numerically shows the influence of the magnetic Prandtl number on the off-diagonal  components of tensor (\ref{alpha5}). The values are taken for mid-latitudes. The $\Pm$ varies by more than one order of magnitude. The numerical values basically grow for growing Prandtl number. Nevertheless, the ratio of the negative quantities  $\langle {u}_x'  \curl_y \vec{B'}\rangle$ and $\langle {u}_x'  \curl_z \vec{B'}\rangle$ remains numerically  always much larger than one, also for the important case of $\Pm<1$.

\begin{figure*}
  \centerline{
   \includegraphics[width=0.75\textwidth]{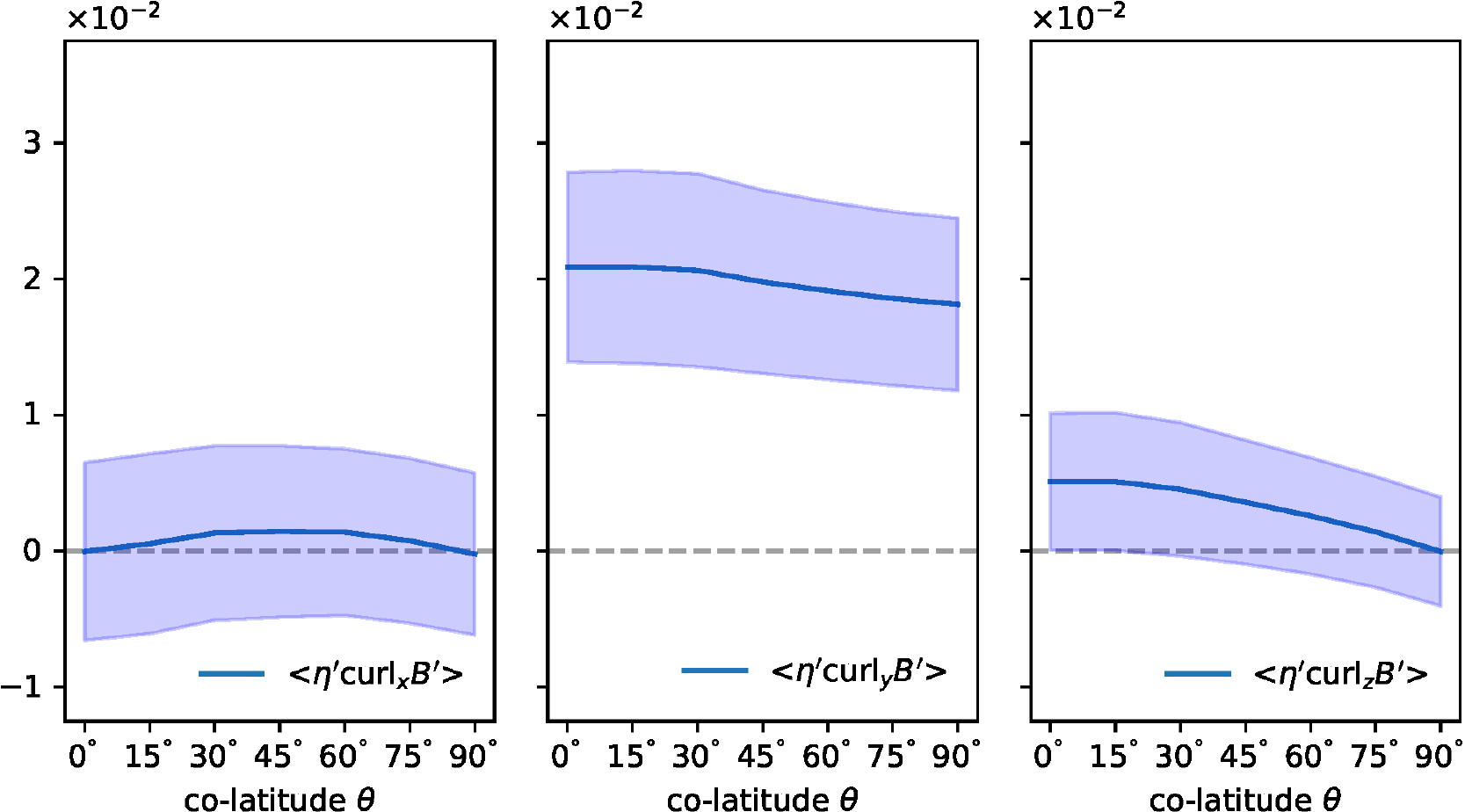}
   }
 \caption{The three  components of the diffusivity-current  vector ${\langle \eta'\curl {\vec B}'  \rangle}/(\urms B_0)$ versus co-latitude. $ \Rm=11$. $\etarms/\bar \eta=0.1$, $B_0=2\times 10^{-8}$,  $\Pm=1$, $\Om=1$.}
\label{fig1}
\end{figure*}

The following numerical simulations in a Cartesian box  with the vertical (radial)  vector  $\vec{g}=(1,0,0) $ have been done with a {\it negative} correlation between diffusivity  fluctuation and vertical velocity, hence $U_x<0$. Again, the applied magnetic field is  azimuthally directed, $\vec{\bar B}= (0,0,B_0)$. We  find
\beg
 {\gamma}= -\frac{\langle \eta'\curl_y {\vec B}'  \rangle}{B_0}, \qquad \alpha_1\Om = \frac{\langle \eta'\curl_z {\vec B}'  \rangle}{B_0}\,.
\label{alphas}
\ende

Figure~\ref{fig1} displays the three components of the diffusivity-current vector as   function of the co-latitude $\theta$. As it should, its radial component   vanishes (left panel). It is also understandable that the advection term $\langle \eta' \curl_y{\vec B}'\rangle$ is positive and does hardly depend on the latitude. According to the first relation in (\ref{alphas}), $\gamma<0$ -- so that the advection velocity ${\vec u}_{\rm adv}$ is directed upwards (i.e., opposite to  $U_x$).

Contrary to this,  the $z$-component of the correlation vector vanishes at the equator -- as it is  expected for a rotation-induced $\alpha$-term. Its maximum is obtained at the poles. According to the second definition (\ref{alphas}),  one finds a positive $\alpha_1$. Note that the negative sign of  the product $\gamma\alpha_1$ is independent of the sign of the correlation of $\eta'$   and $u_x'$.

The simulated components of the correlation vector $\langle \eta'\curl {\vec B}\rangle$ for fixed rotation rate  have been given in Fig.~\ref{fig1}. For a characteristic value $\eta_{\rm rms}/{\bar \eta}=0.1$  of the diffusivity fluctuations,  the  rotation frequency is varied in Fig.~\ref{fig2} to obtain the characteristic  numbers at the northern pole. Obviously, the maximal correlation appears for rotation $\Om\simeq 1$ and will be  suppressed by faster  rotation.

For the ratio (\ref{gammadach}) we generally obtain a value of about five.
The normalised $\alpha$\,effect is
\beg
C_\alpha=\frac{\alpha_1\Om L}{\bar\eta+\etaT}\simeq \frac{3 \alpha_1\Om }{\urms}\frac{L}{\ell}\,,
\label{Calpha}
\ende
with $L$ as the box length in code units. The characteristic turnover time of the turbulence is $\tau_{\rm corr}\simeq \ell/\urms\simeq 2$  in the simulation (also in code units), where $\urms\simeq 0.11$ is set by the amplitude of the forcing. It is $\etaT/\bar\eta\simeq 0.3 \urms^2\tau_{\rm corr}/\bar\eta\simeq 0.3$. According to Fig.~\ref{fig1} and Eq.~(\ref{alphas}), we have $\alpha_1 \Om/\urms\simeq  5\cdot 10^{-3}$ so that
\beg
C_\alpha\simeq 1.5 \cdot 10^{-2}\frac{L}{\ell}\,.
\label{Calpha1}
\ende
The ratio $L/\ell$ gives  the number of cells along the vertical direction,  which obviously must exceed 70 to reach $C_\alpha$ of order unity. This is one of the arguments that it  would  not be easy to simulate such a dynamo in a box.

\begin{figure*}
  \centerline{
   \includegraphics[width=0.75\textwidth]{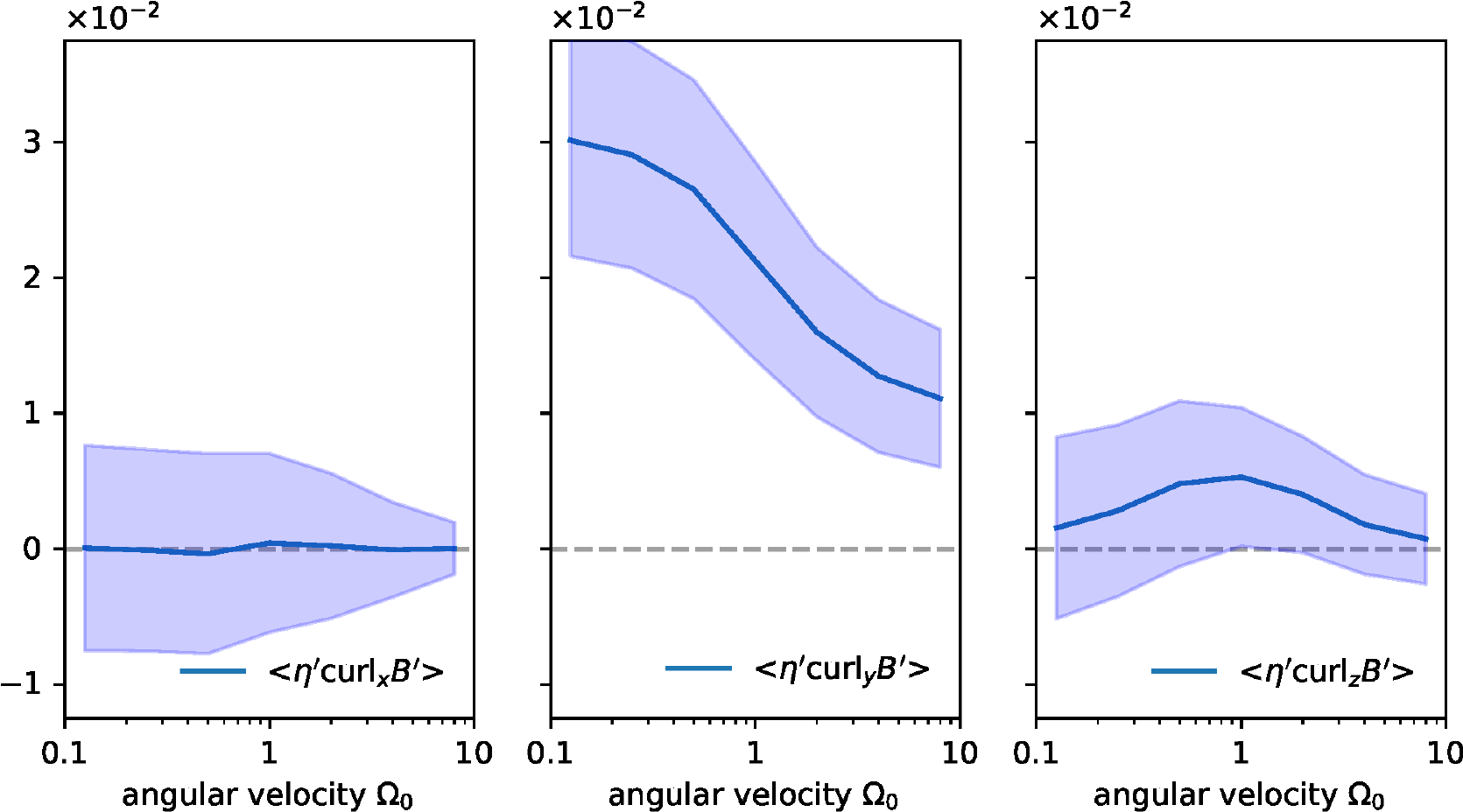}
   }
 \caption{Similar to Fig.~\ref{fig1} but at the northern pole and for increasing rotation rate $\Om$.}
\label{fig2}
\end{figure*}

The dependencies of the diffusivity-current vector components on the rotation rate  $\Om_0$ are shown in Fig.~\ref{fig2}, where for $\Om_0=1$, the rotation period is $2\pi$.  As usual, the  Strouhal number ${\rm St}=\urms \tau_{\rm corr}/\ell$ results of order unity. We also note that  $\Om_0=1$
describes a  rapid rotation with a Coriolis number of $2\tau_{\rm corr}\Om\simeq 4.4$ beyond which the $\alpha$\,effect is strongly quenched by the rotation (Fig.~\ref{fig2}, right panel).  The maximum correlation  exists for $\Om=1$; one cannot increase this value by faster rotation. For $\Om=1$ it is $\hat\gamma\simeq 5$, and this ratio even {\it grows} for slower and/or faster rotation.
A weak rotational quenching can also be observed in the middle panel, where the advection term is reduced (only) by a factor of three when $\Om_0$ grows by two orders of magnitude.

\begin{figure*}
  \centerline{
 \includegraphics[width=0.75\textwidth]{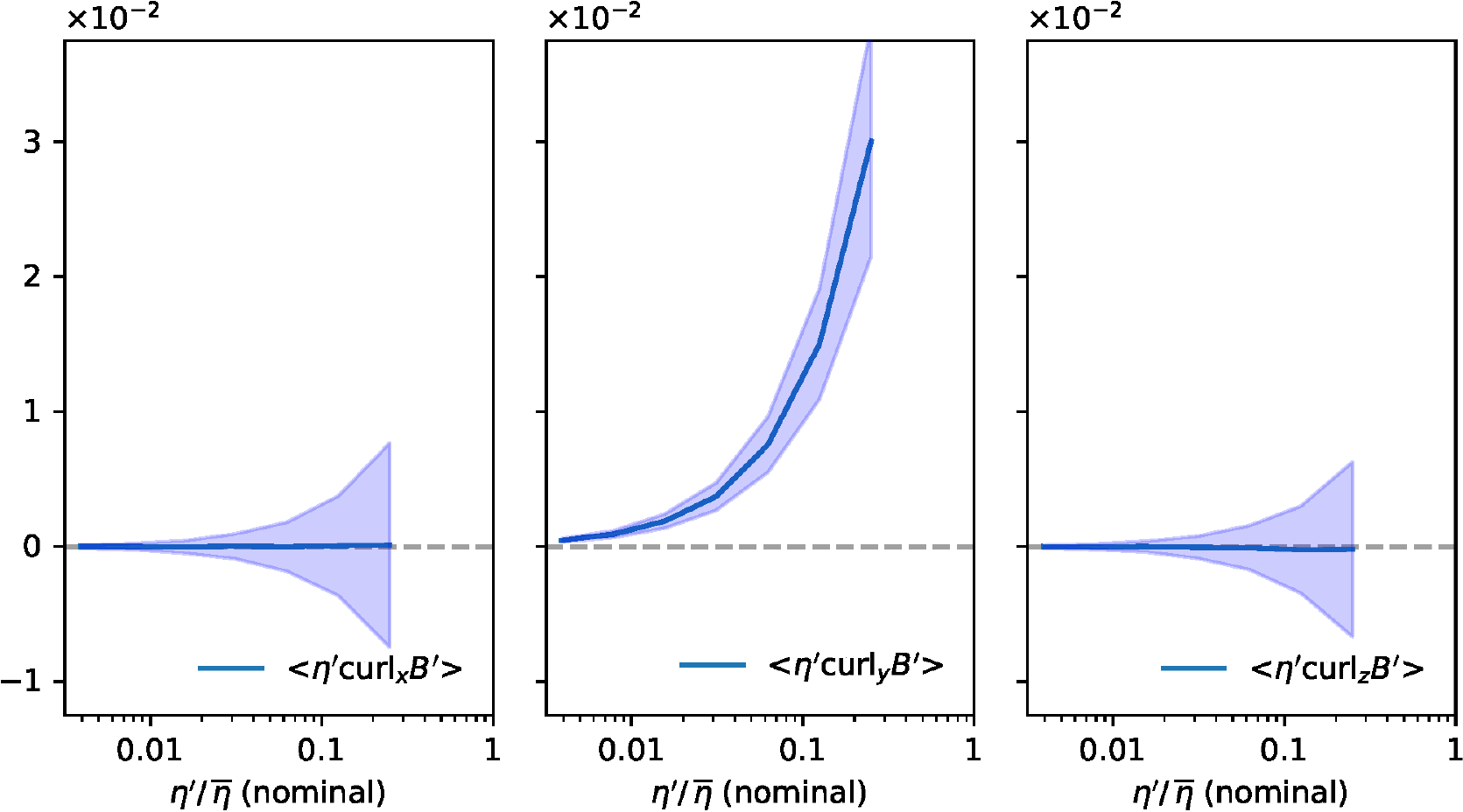}
 }
 \caption{
 The  three components of the vector  $\langle \eta' \curl{\vec B}'\rangle/(\urms B_0)$  for non-rotating turbulence. In accordance with Eq.~(\ref{J2}) only the $y$-component (representing the topological pumping) remains finite. $\Rm=11$, $\Pm=1$, $\Om=0$.}
\label{fig3}
\end{figure*}

Figure \ref{fig3}  refers to non-rotating turbulence with growing ratio of the diffusivity fluctuations, $\etarms/\meta$. As expected, the curve in the middle  panel linearly runs with the normalised diffusivity fluctuation in accordance with the $\gamma$ expression (\ref{gamma}), and it vanishes for $\eta'\rightarrow 0$. For non-rotating turbulence, of course, the two remaining components (including the $\alpha$\,effect) are identical zero -- as shown in the left and the right panel of Fig.~\ref{fig3}. As it should, the advection term plotted in the middle panel also exists for non-rotating turbulent fluids.
We still  have to find out   how the calculated   large values of $\hat\gamma$ influence the  operation of a global dynamo.

% -------------------------------------------------------------------------------

\section{Kinematic ${\bf \alpha^2}$~dynamo models with   field  advection}
\label{models}

% -------------------------------------------------------------------------------

It has been  shown that for rotating turbulence, the above formulated $\alpha$\,effect    is always accompanied by a pumping effect in the direction of the component of the flow vector which is correlated with conductivity fluctuations. For all rotation rates, the ratio $\hat \gamma$ exceeds unity. We now turn our inquiry to the influence of the turbulent field advection  on the operation of an $\alpha^2$\,dynamo. In earlier papers, we already found for disk dynamo models  that a too strong  field advection suppresses the field excitation even under the presence of differential rotation \citep{SE94}.

The geometrically simplest model  is posed by uniform quantities $\alpha$ and $\gamma$ existing
in a gap between two parallel plates embedded in vacuum.
 The vertical distance between the boundaries  is $H$. The eddy diffusivity $\eta_0$ between the plates is assumed as a free parameter, whose actual value is not important for the  result. All quantities are assumed as uniform in the two horizontal directions $y$ and $z$. Then the condition $\div~\vec B=0$ requires that the vertical field $B_x$ does not depend on $x$ hence $B_x=0$ without lost of generality.

  The  equations for this kinematic 1D slab  model may be  formulated with the  normalised quantities
 \beg
  C_\alpha=\frac {\alpha  H}{\eta_0}, \qquad
  C_\gamma =\frac {\gamma H}{\eta_0}
 \label{Cgamma}
 \ende
(let $\Om=1$ for simplicity) so that
\beg
\i\omega B_y
 -\frac{{\rm d}^2  B_y}{{\rm d} x^2}
 = - C_\gamma \frac{{\rm d}  B_y}{{\rm d}x}
 - C_\alpha  \frac{{\rm d}  B_z}
 {{\rm d}x}
 \label{alfadyn1}
 \ende
 and
 \beg
  \i\omega B_z
 -\frac{{\rm d}^2
  B_z}{{\rm d} x^2}=
 - C_\gamma \frac{{\rm d}  B_z}{{\rm d}x} +
 C_\alpha \frac{{\rm d}  B_y}
 {{\rm d}x}\,,
 \label{alfdyn2}
 \\[8pt]
\ende
--- see \citet{P79,M99,RK06}. The real part of the complex frequency  $\omega$ determines the oscillation frequency (in units of the diffusion rate) of a possible dynamo wave along the vertical direction, while the growth rate of the dynamo  is given by the negative value of its imaginary part. We are  mainly interested to know  the critical $C_{\alpha,0}$ for neutral instability, $\Im(\omega)=0$. Let us define the ratio
\beg
\hat\gamma_0=\frac{C_\gamma}{C_{\alpha,0}}
\label{gammahat}
\ende
as describing the influence of the pumping effect on the excitation of kinematic $\alpha^2$\,dynamos.

The vacuum boundary conditions
$
  B_y(0)= B_z(0)=  B_y(1)= B_z(1)=0
$
are applied. For $\gamma=0$ the lowest nontrivial eigenvalue of a stationary solution  is $C_{\alpha,0}=2\pi$. The  solutions  do not depend on the sign of $C_\gamma$ as they do also not depend on the sign of $C_\alpha$.
\begin{figure}
  \centerline{
  \vbox{
 \includegraphics[width=0.45\textwidth]{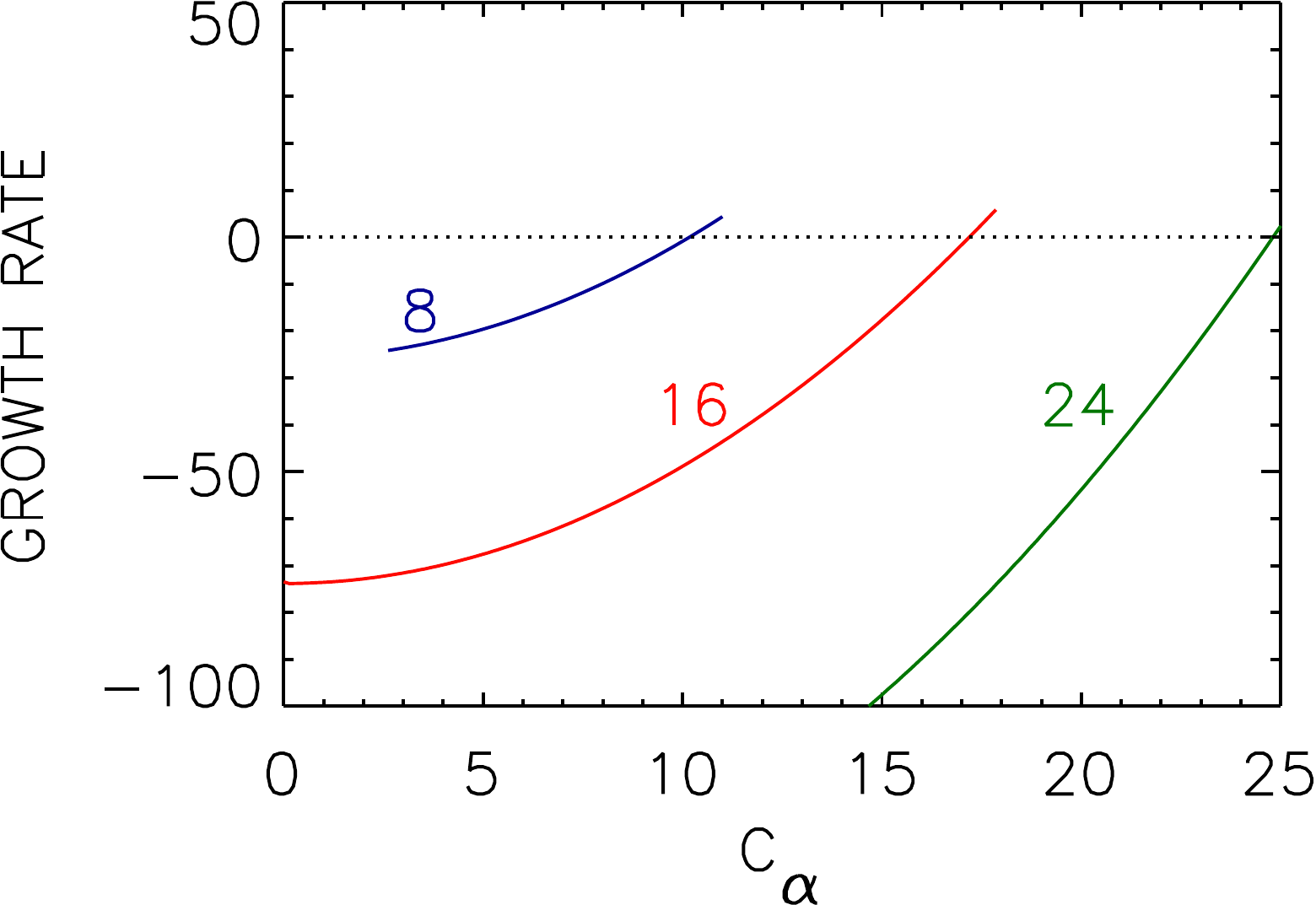}\vspace{6pt}
 \includegraphics[width=0.45\textwidth]{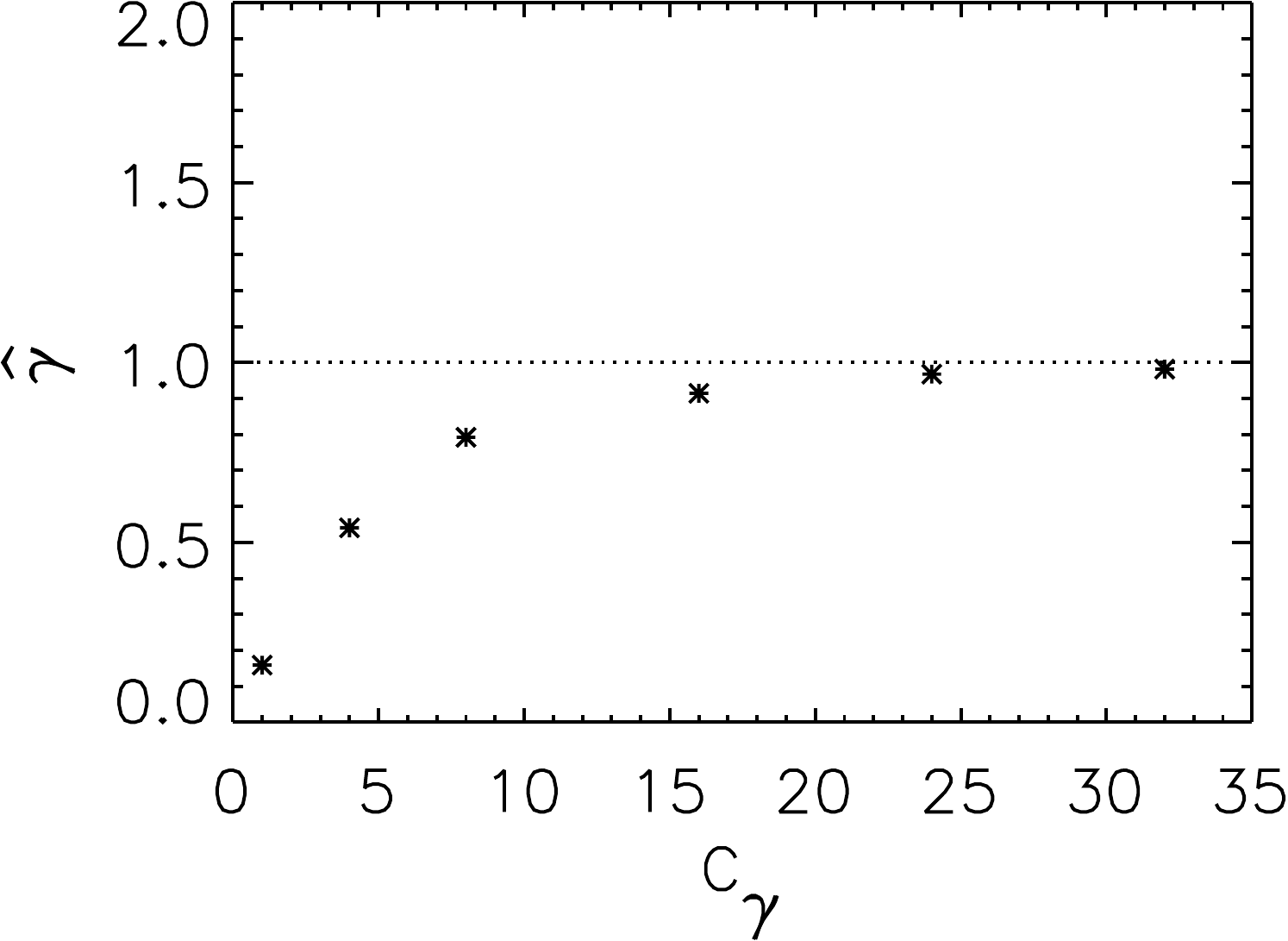}}
 }
 \caption{Upper panel: Growth rates  multiplied with the diffusion time vs. $C_\alpha$   for three plane dynamos with $C_\gamma=8, 16, 24$. All solutions describe waves travelling in vertical direction. Lower panel: The dimensionless ratio $\hat\gamma_0$ versus  $C_\gamma$ for neutral excitation. Kinematic dynamos only exist  as long $\hat\gamma\leq 1$,  the pumping term $C_\gamma$ suppresses the dynamo action.}
\label{figdyn1}
\end{figure}

The upper panel of Fig.~\ref{figdyn1} gives the dynamo's  growth rates for three values of  $C_\gamma$ as function of $C_\alpha$. As usual, for sub-critical  (super-critical) $\alpha$ the modes are decaying (growing), and we find that the  $C_{\alpha,0}$ for neutral instability grows with growing $C_\gamma$. All the critical dynamo solutions for non-vanishing $\gamma$ are oscillating. The lower panel of this figure demonstrates  that for these eigensolutions the ratio $\hat\gamma_0$ does never exceed unity. The 1D  $\alpha^2$\,dynamo, therefore, has no neutral dynamo solution for $C_\gamma>C_{\alpha}$. A too strong radial advection effect  is not compatible with the operation of  $\alpha^2$\,dynamos.  The reason for the suppression   of the dynamo instability  by  dominating  radial advection  is that  the field components  perpendicular to the advection vector are concentrated inwards (or outwards) so that the dynamo domain is reduced and the critical $C_\alpha$ must grow. This destructive action  proves to be even more drastic for $\alpha^2$\,dynamos than for those of $\alpha\Om$-type \citep{BM92,M99}.

Results for a very special spherical  model with $\alpha$\,effect and pumping term are plotted in Fig.~16.10 in \cite{KR80}. The $\alpha$\,effect only  exists in an outer  hemisphere while the diamagnetic  pumping only exists in its inner part. Similar to the above  slab model,  for growing $C_\gamma$ also the critical $C_{\alpha,0}$ grows linearly so that the $\hat\gamma_0$ never exceeds unity. The mode with the lowest dynamo number is a nonaxisymmetric quadrupole with an azimuthally drifting magnetic field.

Because of its relevance  for the concept of conductivity fluctuations, we have designed  a  simple shell-type  dynamo model with an outer turbulence domain filled  with    $\alpha$\,effect independent of the  radius,  and with uniform radial  $\gamma$. The $\alpha$ term is anti-symmetric with respect to the equator. The definitions (\ref{Cgamma}) have been used with the replacement $H\to D$ with $D=(1-\rin)R$ and $R$ the radius of the sphere. The inner boundary is a perfect-conducting one while the outer boundary mimics vacuum, so that the Poynting flux is zero. To illustrate the performance  of the  advection term, examples for  the excited magnetic fields  are plotted in  Fig.~\ref{figdyn2} for a turbulence with outward pumping (top) and inward pumping (bottom). The inner part (or the outer part,  in dependence on the sign of $\gamma$)  of the shell are field-free.  Eigensolutions with dipolar symmetry have the same eigenvalue as those with quadrupolar symmetry. The sign of $C_\gamma$ differs in both models but  without consequences for the excitation condition. For both cases  $|\hat\gamma|=0.8$ is prescribed.  The radial  advection produces  nonaxisymmetric solutions drifting in the azimuthal direction. For $\gamma=0$ the critical eigenvalue for neutral excitation  is $C_{\alpha,0}\simeq 5$, independent of the value of $\rin$ (see Fig. \ref{figdyn2}, middle panel).
\begin{figure}
  \begin{center}
  \includegraphics[width=0.85\linewidth]{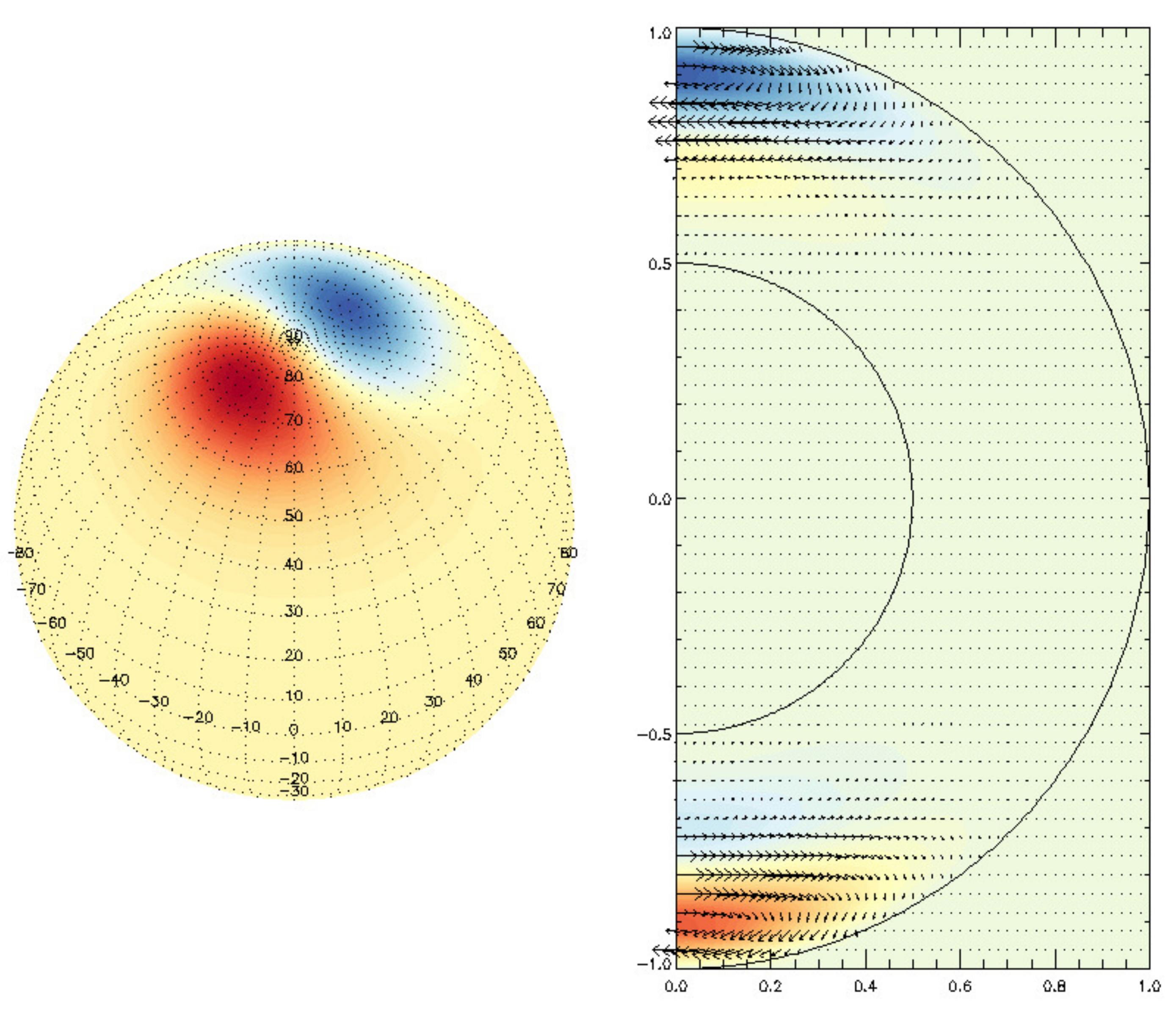}\\
  \includegraphics[width=0.85\linewidth]{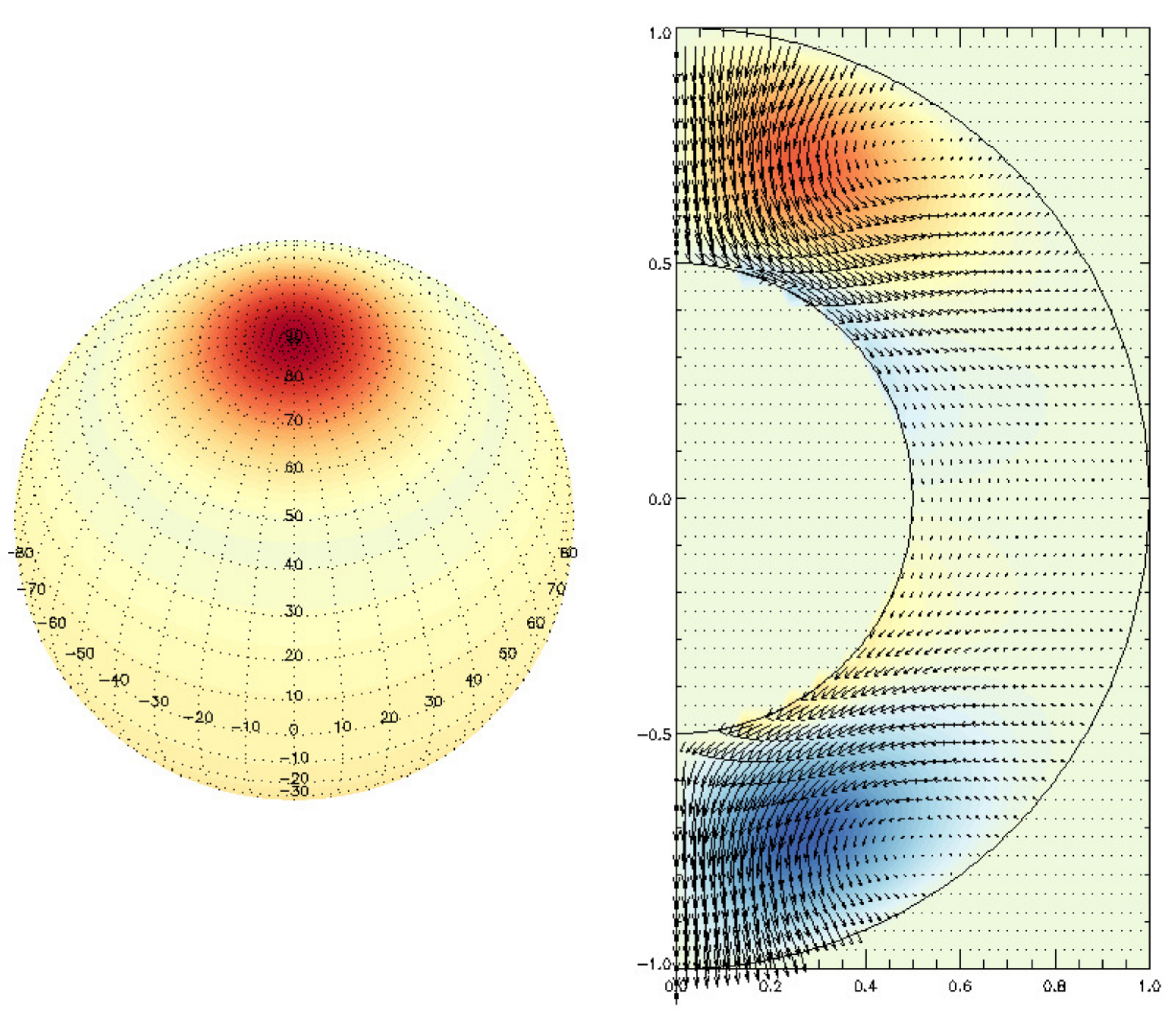}\\
  \includegraphics[width=0.85\linewidth]{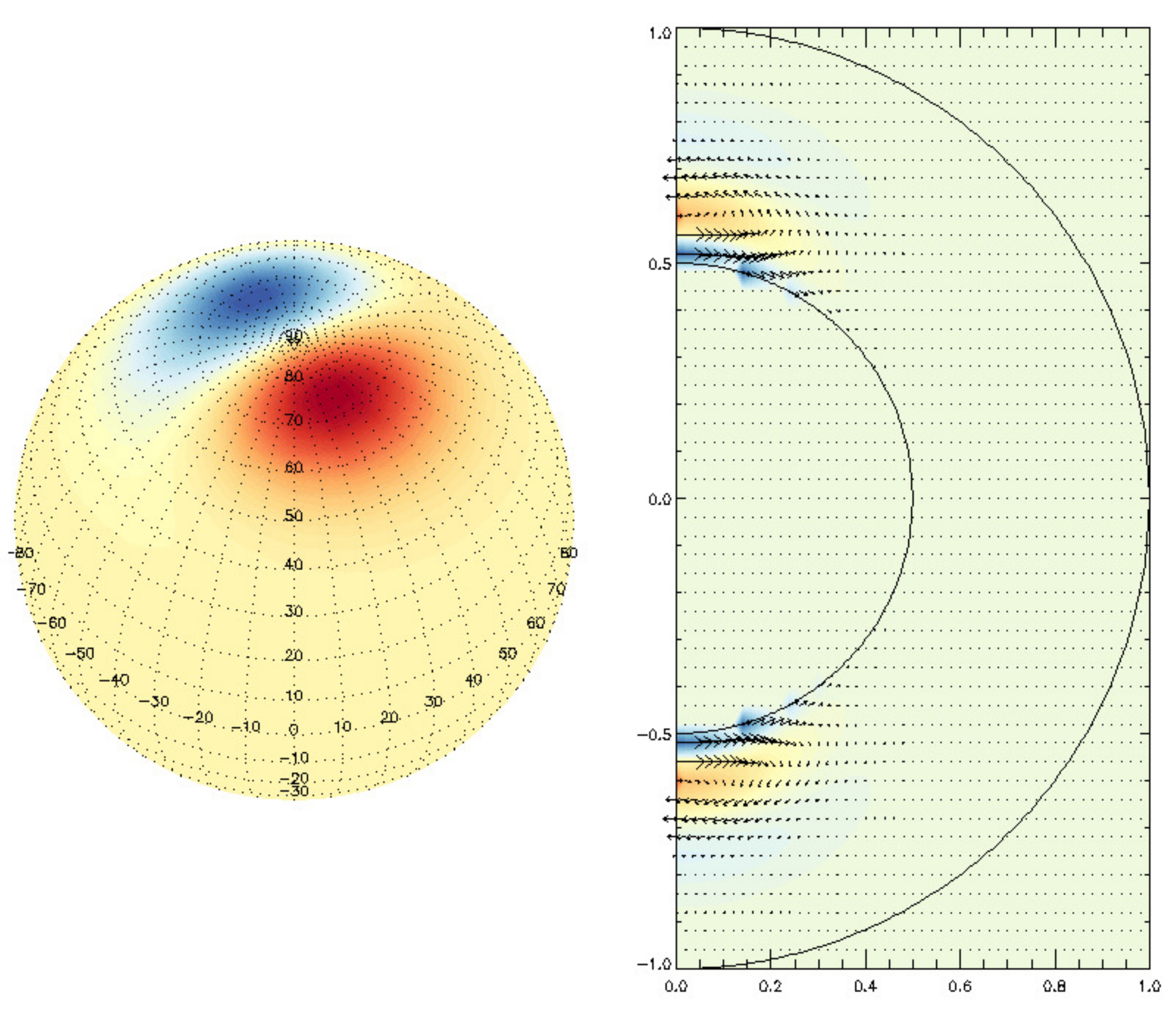}
  \end{center}\vspace{-1ex}
  \caption{Influence of the advection term on $\alpha^2$\,dynamos. The nonaxisymmetric dipolar mode A1 (top) and the quadru\-polar mode S1 (bottom) for $|\hat\gamma_0|=0.8$ are excited by the same value of $C_{\alpha,0}$. The kinematic  axisymmetric $\alpha^2$\,dynamo for $\gamma\!=\!0$ (middle) is shown for reference. The bottom of the turbulence domain is at a $r=0.5$, with a perfect-conducting boundary. The models are embedded in vacuum.}
\label{figdyn2}
\end{figure}

For increasing $\hat \gamma$, the horizontal  field will be more and more concentrated at the inner or the outer boundary (in dependence on the sign of $\gamma$) while the bulk of the shell becomes field-free. The  values of $C_\alpha$ necessary for dynamo excitation grow to unrealistic high values (Fig.~\ref{figdyn3}). A fluid with   values of   $\hat\gamma>1$ and without shear cannot maintain large-scale fields via the $\alpha^2$ mechanism.  For the above calculated  high value of $\hat\gamma\simeq 5$, therefore, kinematic $\alpha^2$\,dynamos are  not possible. With other words, the dynamo only works for
$
C_\alpha\gsim {\rm Max}(5,C_\gamma).
$
 In case that
  $\alpha\simeq \hat\alpha\Om$ (which is true for slow rotation), the dynamo only  operates as long
  as the rotation rate exceeds the critical value of $\Om\simeq \gamma/\hat\alpha$.  The dynamo decays  for $\Om<\Om_1$ where $\Om_1$  denotes the rotation rate where $\hat\gamma=1$. The above mentioned simulations for  solar magneto-convection suggest that indeed $\Om_1\simeq\Om_\odot$.
Figure~\ref{figdyn3} also contains  eigenvalues for an $\alpha^2\Om$~dynamo with the rather flat rotation law $\Om=\Om_0/r^{0.3}$. For the normalised rotation rate $C_\Om=\Om_0 D/\eta_0$ the value $C_\Om=460$ is used. One only finds small deviations from the curves for the $\alpha^2$\,dynamo with $C_\Om=0$. For weak field advection the solutions with the lowest $\alpha_0$ are axisymmetric and oscillating while for stronger pumping the nonaxisymmetric modes prevail which are drifting in the azimuthal direction.

We note that we  only considered the kinematic approximation where any nonlinear feedback of the induced fields onto the turbulence is ignored. In any case, if dynamos ever existed for large values of $\hat\gamma$, they must be rather exotic.

\begin{figure}
  \centerline{
   \includegraphics[width=0.45\textwidth]{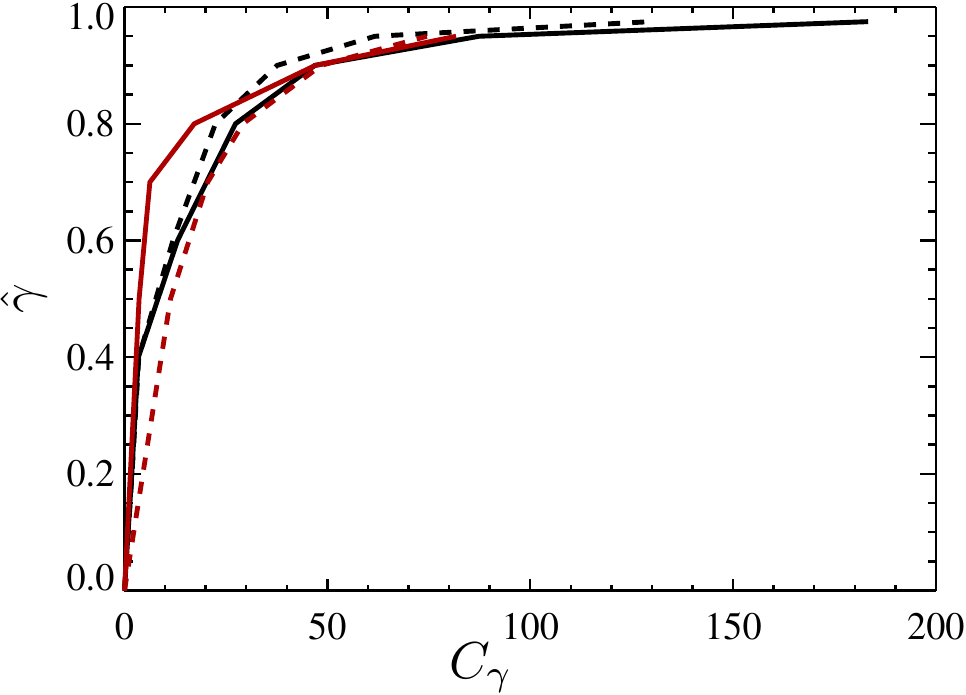}}
 \caption{The values $\hat\gamma_0$ critical for excitation  versus  $C_\gamma$ of  spherical shell dynamo models. The nonaxisymmetric (dashed lines) solutions possess (slightly) smaller $C_\alpha$ than the  axisymmetric solutions (solid line). Dynamo solutions for $\gamma>\alpha_0$ do not exist. $C_\Om=0$ (dark), $C_\Om=460$ (red). The smallest eigenvalue is $C_{\alpha,0 }=5$ for $\gamma=0$. $\rin=0.5$. }
\label{figdyn3}
\end{figure}

% -------------------------------------------------------------------------------

\section{Conclusions}

% -------------------------------------------------------------------------------

If an anisotropy in a conducting turbulent fluid is defined as one that is (only)  in the direction of the conductivity fluctuations, and the velocity fluctuation is correlated, then a turbulent field-advection exists in this direction. It lifts large-scale fields oriented perpendicular to this direction downward or upward, depending on the sign of the correlation.

Our simulations provide the amplitude of this advection term in units of the turbulence velocity. They are on the order of about 10\% of the normalised  resistivity fluctuation  $\etarms/\bar\eta$, while the $\alpha$\,effect is generally smaller.  Its amplitude grows for growing rotation rate until $\Om\simeq 1$ -- declining, however, for faster rotation. On the other hand, the advection term $\gamma$  is numerically almost uninfluenced by the rotation, in accordance with general expectations. As we have also shown that  the $\Pm$-dependence of the results is only weak, one can be sure that in rotating fluids with velocity-correlated conductivity fluctuations, the resulting pumping term $\gamma$ always exceeds the alpha term velocity $\alpha_1\Om$.

As demonstrated in Section~\ref{models}, this  constellation has  severe consequences for associated dynamo models.
There  we have considered two  dynamo models with different geometries. First, a simplified slab dynamo model with two insulating plates and  with a uniform $\alpha$  effect,  including  a vertical turbulence-induced field advection. This model only yields solutions with neutral stability if the $\alpha$ velocity exceeds the advection velocity. The solution for $\gamma=0$ is stationary while otherwise  it forms a vertical dynamo wave. For $\gamma\geq \alpha$, dynamo solutions no longer exist. The results are very similar for spherical shell dynamos. For growing advection effect the most unstable modes become oscillatory but always the dynamos need $C_\alpha>C_\gamma$, i.e. the ratio $\hat\gamma$ never exceeds unity. Pure  $\alpha^2$\,dynamos on the basis of resistivity fluctuations can  thus not work. The same holds for shell dynamos with rather flat rotation laws while the behaviour of $\alpha\Om$~dynamos with  large shear  is still unknown for the case of strong pumping.

% -------------------------------------------------------------------------------

\section*{Acknowledgements}

OG thanks Petri K{\"a}pyl{\"a} for a helpful correspondence. This work used the \textsc{nirvana} code version 3.3, developed by Udo Ziegler at the Leibniz-Institut f{\"u}r Astrophysik Potsdam (AIP). All direct computations were performed on the \texttt{Steno} node at the Danish Center for Supercomputing (DCSC).

%% \subsection*{Author contributions}

%% \subsection*{Financial disclosure}

%% \subsection*{Conflict of interest}

%% \section*{Supporting information}

% -------------------------------------------------------------------------------

\vspace{-3ex}
%%\bibliography{bib/refs}

\end{document}